\documentclass[conference]{IEEEtran}
\IEEEoverridecommandlockouts
\pdfoutput=1
\usepackage{cite}
\usepackage{amsmath,amssymb,amsfonts}
\usepackage{algorithmic}
\usepackage{graphicx}
\usepackage{textcomp}
\usepackage{xcolor}
\def\BibTeX{{\rm B\kern-.05em{\sc i\kern-.025em b}\kern-.08em
    T\kern-.1667em\lower.7ex\hbox{E}\kern-.125emX}}
\begin{document}

\title{Agile Methods in Higher Education: Adapting and Using eduScrum with Real World Projects}
\author{\IEEEauthorblockN{Michael Neumann}
\IEEEauthorblockA{\textit{Dpt. of Business Computing} \\
\textit{Hochschule Hannover} \\
Ricklinger Stadtweg 120 \\
Hannover, Germany \\
michael.neumann@hs-hannover.de}
\and
\IEEEauthorblockN{Lars Baumann}
\IEEEauthorblockA{\textit{Dpt. of Business Computing} \\
\textit{Hochschule Hannover} \\
Ricklinger Stadtweg 120 \\
Hannover, Germany \\
lars.baumann@hs-hannover.de}
}

\maketitle

\begin{abstract}
This Innovative Practice Full Paper presents our learnings of the process to perform a Master of Science class with eduScrum integrating real world problems as projects. We prepared, performed, and evaluated an agile educational concept for the new Master of Science program \textit{Digital Transformation} organized and provided by the department of business computing at the University of Applied Sciences and Arts - Hochschule Hannover in Germany. The course deals with innovative methodologies of agile project management and is attended by 25 students. We performed the class due the summer term in 2019 and 2020 as a teaching pair. The eduScrum method has been used in different educational contexts, including higher education. During the approach preparation, we decided to use challenges, problems, or questions from the industry. Thus, we acquired four companies and prepared in coordination with them dedicated project descriptions. Each project description was refined in the form of a backlog (list of requirements). We divided the class into four eduScrum teams, one team for each project. The subdivision of the class was done randomly. Since we wanted to integrate realistic projects into industry partners' implementation, we decided to adapt the eduScrum approach. The eduScrum teams were challenged with different projects, e.g., analyzing a  dedicated phenomenon in a real project or creating a theoretical model for a company's new project management approach. We present our experiences of the whole process to prepare, perform and evaluate an agile educational approach combined with projects from practice. We found, that the students value the agile method using real world problems. However, the results are mainly based on the summer term 2019, this paper also includes our learnings from virtual distance teaching during the Covid 19 pandemic in summer term 2020. The paper contributes to the distribution of methods for higher education teaching in the classroom and distance learning.
\end{abstract}



\section{Introduction}
Agile methods such as Scrum \cite{Schwaber.2021} or Extreme Programming (XP) \cite{Beck.2000} are established approaches in software development and continue to show increasing usage in companies worldwide \cite{VersionOne.2020}. The motivation to use agile methods is often argued due to the increased dynamics of requirements in software development projects or the need to be able to deliver product increments to users and customers \cite{Williams.2010}. Agile methods are characterized by their adaptability, the collaborative, and incremental and iterative approach \cite{Abrahamsson.2002}. The agile manifesto describes a definition of values and principles for agile methods \cite{Beck.2019}.

The increasing usage of agile methods in practice has led to more and more integration into teaching in the fields of engineering, computer science, or information systems. Lecturers adapted agile methods such as Scrum or XP for the educational context and used them in projects with students or lab courses (e.g., \cite{Elgrably.2020, Otero.2020}). In addition to applying methods that are of high relevance for the student's professional future, agile methods offer further advantages in teaching \cite{Lundqvist.2019}. Mainly, the students work in teams, which means that one can expect an improvement in social skills. In addition to the necessary communication and collaboration in agile practices, these also include the processing of feedback from customers and other stakeholders. Furthermore, it is common for the teams to present their work results on a regular base, thereby training these skills as well. The integration of agile methods also offers the opportunity to integrate real-world problems and challenges into teaching \cite{Jennings.2019, Mahnic.2015}. This approach comes very close to project and/or problem-based learning, although not all specific activities of these didactic methods have to be taken into account \cite{Lang.2017}. 

As shown in Section \ref{s2:rel_work}, several studies are aiming to analyze how agile methods can be integrated in higher educational contexts. We also present an overview of literature, which is dealing with eduScrum. We found out that the use of eduScrum in higher education has not been extensively investigated in research yet. This paper addresses this research gap by combining practical, project-based teaching with eduScrum and leads to the following research questions:
\begin{itemize}
    \item RQ1: How can eduScrum be used with real and practical problems in higher education? 
    \item RQ2: How do the students value the work with eduScrum?
\end{itemize}
We have adapted the eduScrum method for integration in a (Master of Science) project management course and combined our approach with projects from practice partners (companies). The results of our study include the summer term of 2019. 

The paper at hand is structured as follows: In Section \ref{s2:rel_work}, we present related work. We explain the research approach in Section \ref{s3:res_design}. In the subsections, we describe the course information, our adapted eduScrum approach, the class organization and the data collection and analysis. We present the results and the answers of our research questions in Section \ref{sec4:results}. \\We describe the Limitations in Section \ref{s5:lim} before the paper closes with a summary in Section \ref{s6:conc}.

\section{Related work}
\label{s2:rel_work}

In the literature, several approaches for the integration of agile methods into the various forms of teaching are described. We found also several adaptions of the agile manifesto for the educational context, which consider the peculiarities in knowledge transfer and teaching and formulate corresponding values (e.g., \cite{Kamat.2012, Krehbiel.2017, Madhuri.2018, Stewart.2009}). With the increasing spread of agile methods in practice, more and more researchers and lecturers were interested in using agile methods in an educational context. The increasing interest is shown by several studies (e.g., \cite{Otero.2020, Salza.2019}). According to Otero et al., models such as Scrum or Extreme Programming are being adapted in university teaching and, in particular, the area of computer science \cite{Otero.2020}. Mahnic shows in his literature overview the different facets of Scrum in educational contexts and how the agile method is applied in the classroom in order to be able to include real world problems and challenges \cite{Mahnic.2015}. Salza et al. present in their systeamtic literature review that several agile methods are used in higher education in different teaching formats, like student projects, lab classes or courses \cite{Salza.2019}. 

Kropp and Meier describe their experiences with an agile approach to organizing and performing a software engineering (SE) course \cite{Kropp.2013}. The authors use various agile methods (e.g., Scrum and XP) for imparting knowledge and project work in the class. Also, Linos et al. used Scrum for organizing their computer science and software engineering class and integrated IT professionals to their educational design \cite{Linos.2020}. The authors aimed to provide a real world situation in the class. Based on the survey results the students value the form of organization and idea to involve IT professionals. Mahnic show his observations and the students perception of the integration of Scrum in a SE course \cite{Mahnic.2010}. Scharf and Koch present similar results in their practical report on the use of Scrum in a SE course \cite{Scharf.2013}. According to Scroeder et al., Scrum is also used in software development lab courses \cite{Schroeder.2012}. The authors emphasize the focus on product creation and continuous improvements instead of the completeness of the product. 

In addition to the specific adaptations of Scrum, XP, and other agile methods for the educational context, agile methodologies for use in teaching were also presented in the past. A well-known example of this is eduScrum, which was published by Wijnands et al. and is described in an eduScrum Guide \cite{Wijnands.2020}. Eduscrum was initially adapted for use in Dutch schools and is now increasingly used in higher education (e.g., \cite{Cardoso.2018, Tudevdagva.2020}). The integration of eduScrum has not yet been investigated as much as other agile education approaches. Ferreira and Dias Canedo present in their study the integration of eduScrum in a large Informatics study programm. They found, that the students were more motivated in lab courses. Dinis-Carvalho et al. evaluate eduScrum and lean teaching. The authors present a guideline for the use of the both methods that they have created based on best practices \cite{Dinis.2017}.

The presented studies are investigating the use of eduScrum in higher education, but do not put a mainly focus on practical aspects, e.g., in form of integration a project based learning approach or an integration of real world problems from practice partners and their stakeholder. Thus, we decided to create a study with the aim to provide results of our eduScrum approach to the community.
\section{Research design}
\label{s3:res_design}
\subsection{Course information}
The Master of Science program \textit{Digital Transformation} is supervised and offered by the Business Computing department since the winter term 2018/2019. A student group in the master program comprises 25 persons. The course \textit{Innovative Methods of Project Management} is offered annually in the summer term. We are an onsite university, so the courses are designed for classroom teaching in person. Since the Covid 19 pandemic reached Germany in March 2020, the course has been offered exclusively via virtual distance learning. Thus, we performed the course once permanently in attendance (summer term 2019) and once virtually via distance learning (summer term 2020).

In the course, the students should get to know and understand new methods and challenges of project management. They should be able to handle them based on systematic approaches and methods. We present the learning objectives in Table \ref{tab-1}.

\begin{table}[htbp]
\caption{Learning objectives}
\begin{center}
\begin{tabular}{p{5cm}}
\hline
\textbf{Learning objectives}\\
\hline
Understand the differences and characteristics of agile, plan-based (phase-oriented) and hybrid methods\\
\hline
Understand the challenges of intercultural project teams\\
\hline
Leadership and team coordination for different project sizes \\
\hline
Conflict management in projects\\
\hline
Presentation of status reports for selected stakeholder \\
\hline
Personnel management and motivation of virtual distributed international teams \\
\hline
\end{tabular}
\label{tab-1}
\end{center}
\end{table}

Each term is divided into different phases: lecture period (16 weeks), examination period (three weeks) and lecture-free period (seven weeks). The course is planned with total effort of 180 hours (68 hours in attendance, 112 hours self study). The 68 hours are planned as lecture units, as well as during the Covid 19 pandemic. 

\subsection{Adapted eduScrum approach}
As mentioned in Section \ref{s2:rel_work}, eduScrum was developed in the Netherlands for lessons in schools \cite{Wijnands.2019}. Weijnands et al. created an eduScrum Guide, which describes agile practices, roles, artifacts and rules as well as values and principles for executing eduScrum \cite{Wijnands.2020}. As the eduScrum Guide was developed for the school system and not for a combination with external stakeholder, which provides projects and exercises, we decided to adapt eduScrum to our needs. 

The eduScrum Guide describes three roles: Product Owner (teacher/lecturer), Students team and a eduScrum Master (like a team captain). We adapted the role concept of eduScrum; we added an Agile Coach role and changed the Product Owner responsibilities and tasks. Thus, we defined an eduScrum team consisting of four roles:
\begin{itemize}
    \item \textit{Product Owner:} Based on Scrum \cite{Schwaber.2021}, we have assigned the role of the product owner to the stakeholder (professional) of the practice partner (company). He is responsible for the requirements of the student team and thus also provides the product backlog (prioritized list of requirements). The role also acts as a contact person for questions regarding the content of the students.
    \item \textit{Agile Coach:} The lecturers take on a coaching role in the adapted eduScrum framework. The agile coach role supports the students in questions relating to the eduScrum approach and/or content-related questions relating to the underlying theory of the course. In addition to specialist knowledge, this also includes further areas of knowledge such as research methods or the organization of the class.
    \item \textit{eduScrum Master:} Each student team chooses an eduScrum Master. The assignment of the role can be carried out on a rolling basis in each sprint. Thus, every student has the opportunity to take on the role once during the term. The eduScrum Master coordinates the organization in the student group and acts as an impediment remover. If obstacles are identified, it is the responsibility of that role to remove the obstacles. Obstacles can affect both technical restrictions (e.g., concerning tools) and content-related coordination with the product owner.
    \item \textit{Student´s team:} The student's team acts as a self-organizing team. The team decides how to implement the requirements and is responsible for structuring the work tasks in the individual sprint. We also assume that all the necessary technical skills for processing the tasks are available in the team, i.e., an interdisciplinary team. The size of the student´s team is restricted to three to seven people in order to ensure, that the team is able to coordinate and split up the work during a sprint. The students are assigned to the same team throughout the term.  Changes or new team formations are not planned. 
\end{itemize}

In addition to the role concept, we adapted the events (meetings) to our needs and those of the stakeholders of our practice partners:

\begin{itemize}
    \item \textit{Sprint Planning:} The primary focus of this event is to plan and organize the work for the current sprint. In coordination with the product owner, requirements (product backlog items) are selected for the sprint and, if necessary, ambiguities clarified. The selected product backlog items (PBI) represent the sprint goal and are the basis for the student team to create the work steps for achieving the task. It is also part of the sprint planning to choose an eduScrum master. This choice is made by the student team. It is only omitted if a team agrees that the role of the eduScrum Master should not be filled again. Whether a student team estimates the tasks in terms of complexity or effort, for example, is up to the team. If a team decides to estimate its efforts, the agile coach helps learning and performing an estimation method (e.g., Planning Poker \cite{Mahnic.2012} as described according to the eduScrum framework \cite{Wijnands.2020}). The Sprint planning event is time boxed to 45 minutes.
    \item \textit{Sprint Review:} In contrast to the eduScrum Guide \cite{Wijnands.2020}, we plan a sprint review at the end of every sprint. We argue this adaptation with the shorter length of the sprint and the advantages of regular feedback on the sprint transition. In the sprint review, the student team presents the results of the individual sprint to the product owner and agile coach. The product owner has the option of accepting the content of these results but also refusing acceptance. In this case, the PBIs that have not been accepted must be transferred back to the Product Backlog to be taken into account for the coming sprints. When it comes to feedback, the agile coach focuses on the learning objectives and the scientific approach to developing the increments. The duration of the sprint review is set at 30 minutes.
    \item \textit{Sprint Retrospective:} We did not integrate the event as a mandatory element in our eduScrum approach. It is up to the student teams if they want to conduct a retrospective for a sprint. The agile coach supports the desire to conduct a retrospective and acts as a facilitator who also shows the team various implementation methods, if the student team requests the support. From our perspective, a retrospective should not be longer than 45 minutes.
    \item \textit{Stand Up Meeting:} The eduScrum Guide recommends a five-minute stand-up meeting at the beginning of each lesson so that the team can synchronize. We follow this recommendation as far as possible and add the individual meetings organized by the student team. The transparency about their approach, the processing status, and any impediments or problems is valuable so that the student team has the opportunity to react to it.
    \item \textit{Refinement:} Refinement always takes place in the week when no sprint planning occurs. The event is time boxed to half an hour and is used mainly for the product owner to present and explain new or changed PBIs. The student team can (and will) discuss open questions about individual tasks with the product owner. If the student team estimates the complexity or effort involved in sprint planning, this can also be done in refinement if there is any time.
\end{itemize}

Next, we adapted the artifacts of eduScrum. For example, we did not consider all the described artifacts from the eduScrum Guide to our approach like the \textit{definition of fun} or changed existing artifacts such as the \textit{flap}: 

\begin{itemize}
    \item \textit{Product backlog:} The eduScrum framework provides no artifact, which is usually named backlog in the terminology of agile methods (e.g., \cite{Abrahamsson.2002, Williams.2010}). The product backlog contains items that represent specific requirements. These requirements can be specified and documented in different ways. A well-known method (which is also described in the eduScrum Guide \cite{Wijnands.2020}) is user stories \cite{Cohn.2004}. Concerning any quality or acceptance criteria, we follow the eduScrum Guide and recommend that the student teams and product owners use Definitions of Done. The Product Owner is responsible for the Product Backlog. It decides on the prioritization in the backlog, i.e., the order of the entries based on their importance (top-down approach). However, the student team (and the agile coach) can add their requirements to the product backlog.
    \item \textit{Sprint backlog:} EduScrum describes the Run Up Chart \cite{Wijnands.2020} as an artifact for the measurement and transparency of the learning progress. We decided to provide the student teams with an agile practice from Scrum: The Sprint backlog. The Sprint backlog contains the selected PBIs for the sprint and the tasks are belonging to each PBI. The student team decides how the sprint backlog is visualized. The teams used various tools for this (see section \ref{sec4:results}). The sprint backlog always shows the current processing status and should be maintained accordingly by the student team. It is also used for communication with  the product owner and agile coach during the sprint.
    \item \textit{Definition of Done:} The definition of done (DoD) is a well-known artifact used in Scrum \cite{Schwaber.2021} and other agile methods. The student team use the definition of done to describe acceptance criteria or quality criteria for specific requirements. Also, the DoD can be used to define \textit{work agreements} \cite{Wijnands.2020} or organizational facets of the teams work. 
    \item \textit{Product increment:} The artifact represents the result of a sprint. The increment consists of different partial results. These can be the acquisition of knowledge as well as concrete work results such as model drafts. 
\end{itemize}

Finally, we have also adjusted the length of the sprint. The eduScrum Guide defines a sprint length of \textit{approximately seven weeks} \cite{Wijnands.2020}. Our impression was, that the iteration length would be too long for the coordination and delivery of product increments and the assessment. Because of a shorter coordination between the students teams and the product owner, we defined a sprint duration of two weeks. The meetings described above (planning, review, retrospective) take place for the transition between two sprints. At the weeks with no sprint transition, a refinement is carried out with the aim of maintaining the product backlog, preparing the following planning meeting and clarifying questions or ambiguities. A sprint can only be canceled by the roles of Agile Coach or Product Owner.

However, the eduScrum framework is designed for educational contexts; we identified several needs to adapt the approach. We increasingly used the guidelines of agile methods from practice such as Scrum \cite{Schwaber.2021} for the adaption because we aimed to integrate well-known practices and artifacts to enable students to gain practical experience. This integration also facilitates agile coaches and the students cooperation with the product owners, which should have a lot of experience with many of these practices and artifacts from their daily business.

\subsection{Course organization}
As mentioned above, the course \textit{Innovative Methods of Project Management} is planned once a year in the summer term. We prepared and performed the classes as a teaching pair during the summer terms of 2019 and 2020. The course was planned with four lecturing units with 45 minutes each per week and in presence. The form of examination is a seminar paper consisting of two components: a written paper and a presentation. Both components make up 50 percent of the grade.

\paragraph{Preparation and planning tasks}
We asked seven companies for their support to provide a product owner and a project in summer term 2019. All companies take part in our research and educational network. Finally, we selected four companies based on their relevance to the learning objectives of the course (see Table \ref{tab-2}). We have anonymized the names of the companies for reasons of confidentiality.

\begin{table}[htbp]
\caption{Overview of the companies and projects in summer term 2019}
\begin{center}
\begin{tabular}{l|l|p{5cm}}
\hline
\textbf{Company name} & {Industry} & {Project goal} \\
\hline
Comp. Lannister & Banking & Analyze the status quo of project management methods in use and identify optimization potential \\
\hline
Comp. Stark & Insurance & Identify success factors of agile transition and develop an assessment method for an agile transition \\
\hline
Comp. Targaryen & Energy & Analysis of project management methods in use with the aim to develop a hybrid (tailorable) model \\
\hline
Comp. Mormont & Chemicals & Evaluation of hybrid project management models \\
\hline
\end{tabular}
\label{tab-2}
\end{center}
\end{table}

The selected projects are based on real world problems and challenges. The product owner from each company are responsible for preparing a product backlog and maintaining it with the students team.

\begin{figure}[htbp]
\centerline{\includegraphics[scale=0.26]{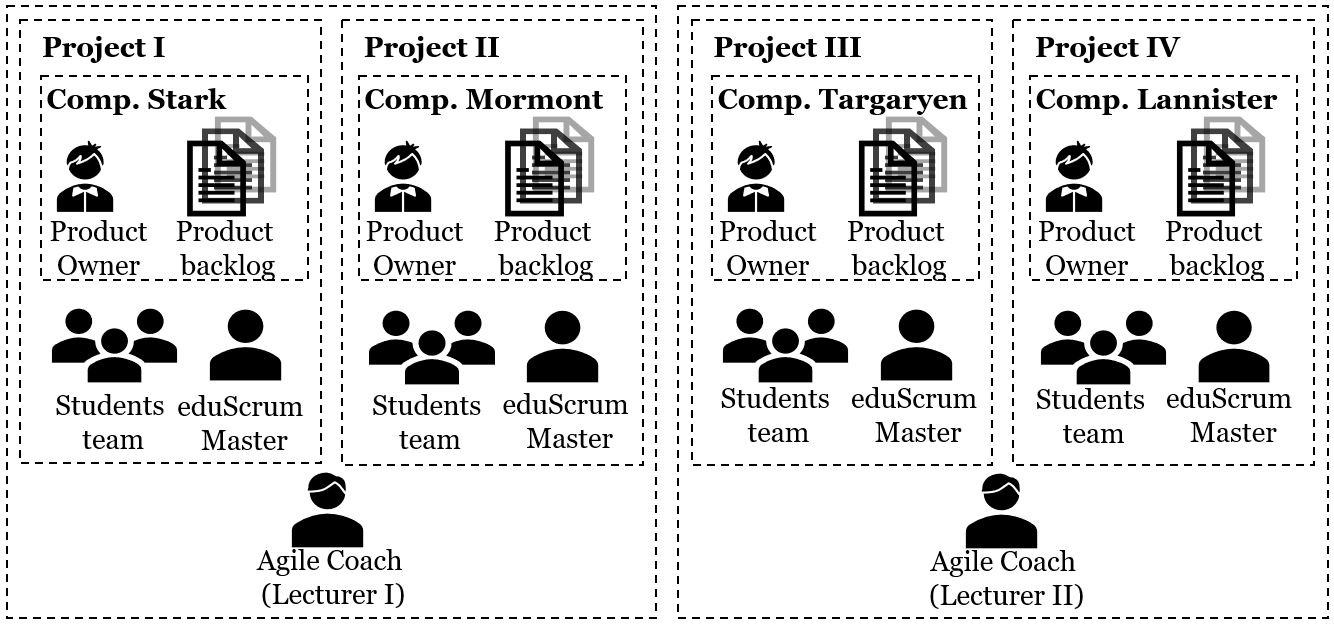}}
\caption{EduScrum class setup}
\label{fig1}
\end{figure}

We divided the student group (25 people) and assigned the students randomly to four subgroups and the associated projects (see Figure \ref{fig1}).

\paragraph{Instruction events and performing eduScrum}
Before we started the iterative approach according to eduScrum, we held an introductory event, which aimed to deal with the organizational aspects of the course and a repetition of the relevant content. For this purpose, we introduced ourselves and the course (e.g., objectives, content, type and rules of examination, subgroups assignment) and asked the students about their knowledge of project management. The repetition of the content includes the fundamentals of agile, plan-based, and hybrid project management (e.g., the agile manifesto, Scrum waterfall model). 

We defined the second event as the kick-off of the student teams with the product owner and planned it as four individual events. We split up into two student teams and took part in the meetings. The kick-off event aimed to ensure a consistent understanding of the project content and goals in the student teams and to plan the organizational procedure in a sprint rhythm.

\begin{figure}[htbp]
\centerline{\includegraphics[scale=0.32]{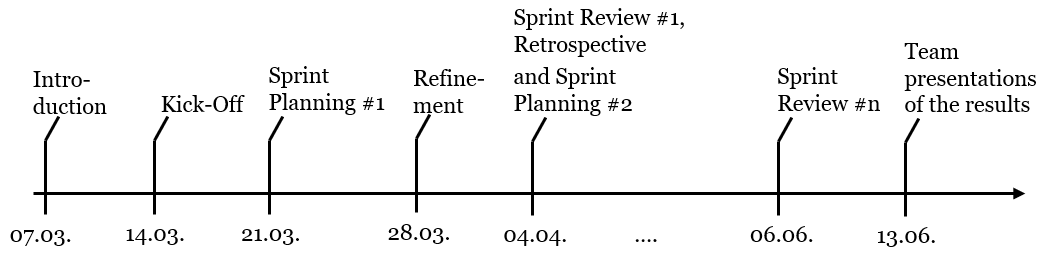}}
\caption{Timeline summer term 2019}
\label{fig2}
\end{figure}

After the kick-off event, we switched to the iterative approach according to eduScrum (see Figure \ref{fig2}). As explained above, the sprints are planned to be two weeks in length. We have planned the appointments (sprint planning, retrospective, review, refinement) so that an agile coach can participate in each appointment. The appointments are scheduled in the course attendance times. The product owners and agile coaches took part in the weekly events and gave feedback on the increment results or provided support with questions and ambiguities. Even though the incremental results of the students teams were not taken into account for grading, they benefit from peer assessments \cite{Fagerholm.2013} each Sprint due to the feedback they got from the Product Owner and Agile Coach. The product owner always explain the current product backlog and the content of the new tasks.

\subsection{Data collection and extraction}
We used a survey to evaluate our teaching quality. The survey questionnaire consists of two categories of question types: closed and open questions. The closed questions are divided into five sets of questions (for example, on the workload and assessment of the lecturer). The open questions cover four areas in which the students can, make suggestions for improving the course of the room furnishings. The data evaluation of the questionnaire was carried out automatically. A draft of the survey can be found in Appendix A.

In addition to the survey, we collected data, e.g. due to observe the eduScrum events. As part of the iterative eduScrum approach, we held informal discussions with both, student teams and product owners, and document them in protocols. According to Yin, this is an appropriate method when extensive data collection methods such as semi-structured interviews are not possible~\cite{Yin.2009}. During the term it was not possible to conduct semi-structured interviews due to time constraints. Also, we wanted to avoid to distract the students. In addition to the informal discussions, we held review discussions with the product owners at the end of the summer term. We extracted and analyzed the data with Excel and used the extracted data to validate the survey results.

Finally, we considered the exam documents (paper and presentations) of the student teams.

\section{Results}
\label{sec4:results}
\subsection{Overview of the results}

The survey was answered by 21 students. The evaluation was carried out after the presentation of the results by the student teams to minimize the risk of bias.

During the term, the student teams used various tools to share documents, coordinate their work and hold meetings. An overview of these tools is given in Table \ref{tab-3}. Many teams used Dropbox to exchange documents. All teams used Microsoft Office applications such as Word, Excel, or PowerPoint to prepare the examination results. Trello was chosen for the visualization of the sprint backlogs. From today's perspective, it is certainly interesting that the student teams already used Microsoft teams for communication in the distributed form of work in the summer of 2019.

\begin{table}[htbp]
\caption{Overview of the used tools}
\begin{center}
\begin{tabular}{l|l}
\hline
\textbf{Tool} & \textbf{Objective of use} \\
\hline
Dropbox & Sharing and management of documents \\
\hline
MS Excel & Documentation  \\
\hline
MS Powerpoint & Results presentation  \\
\hline
MS Teams & Team communication and coordination  \\
\hline
MS Word & Documentation  \\
\hline
Skype & Communication and Meeting management  \\
\hline
Trello & Visualization of the sprint backlog \\
\hline
WhatsApp & Team communication and coordination \\
\hline
\end{tabular}
\label{tab-3}
\end{center}
\end{table}

As described in Section \ref{s3:res_design}, in addition to the quantitative questionnaire, we held various informal conversations, mostly during breaks or before or after the face-to-face events, like Sprint Plannings. The discussions were held with all teams and Product Owners.

\subsection{Answering the Research Questions}
We primarily use the quantitative questionnaire to answer the research questions. In this, the students have the opportunity to answer the closed questions based on a Likert scale with the values 1 (completely applicable) to 5 (completely inapplicable).

Our \textbf{first research question} is: \textit{How  can  eduScrum  be  used  with  real  and  practical problems in higher education?}

We present our adapted eduScrum approach and the reasons for the adaptations in Section \ref{s3:res_design}. 

While using eduScrum with real projects from companies, we identified the risk that the students will lack the detailed theoretical knowledge transfer. The students invalidated this assumption with their evaluations (see Table \ref{tab-4}). We have also assumed that visibility and responsiveness will decrease both during and outside of the eduScrum event through our coaching mentoring role. The students also used this negative assumption differently when evaluating the corresponding statements in the survey. The teachers' assessment was consistently positive, with an average of 1.3 or better. This also includes responsiveness, for example, through quick response to questions. From our perspective, the positive evaluation and the associated appreciation for our commitment are particularly gratifying. It shows that the students know how to assess how time-consuming preparations can be, such as establishing contact, selecting, and coordinating with practice partners for such a teaching format. This aspect has also been confirmed by various informal discussions in which the students (and the product owners) repeatedly gave positive feedback on the effort on our part.

\begin{table}[htbp]
\caption{Average scores on the course content and lecturers quality}
\begin{center}
\begin{tabular}{p{5cm}|p{2cm}}
\hline
\textbf{Survey item} & \textbf{Average score} \\
\hline
Overall, I rate the content of the course positive. & 1.7 \\
\hline
The course content was conveyed clearly. & 1.8 \\
\hline
The lecturer's engagement was high.  & 1.3 \\
\hline
The lecturer responded appropriately to questions. & 1.2 \\
\hline
The lecturer was outside of the class available for support. & 1.2 \\
\hline
My personal impression of the lecturer's occurrence was good. & 1,1 \\
\hline
Overall, I rate the lecturer positively. & 1.1 \\
\hline
\end{tabular}
\label{tab-4}
\end{center}
\end{table}

The \textbf{second research question} that we want to answer with this paper is: \textit{How do the students value the work with eduScrum?}

From our perspective, a decisive aspect in the evaluation of eduScrum by the students is the students' understanding of the connection between real problems and challenges. We focus on two survey items from the questionnaire (see Table \ref{tab-5}):
\begin{itemize}
    \item \textit{My understanding of real world problems related to project management was promoted.}
    \item \textit{The importance of the material for my professional activity became clear to me.}
\end{itemize}
Both items were rated as applicable with a value of 1.5 or 1.4. This results show that we achieved one of our main goals in the class in summer term of 2019: The integration of real-world problems into the course and thus into the training of the students. This is also shown by various comments from the students in the survey, such as: "\textit{Working with companies was better than a lecture.}" or "\textit{Gather very practical and good experience.}"

It is also of interest to us that the students rate eduScrum positively as a didactic method (see the last score in Table \ref{tab-5}). This aspect is also confirmed our conversations with the students. 

However, some students also comment in the survey the improvement option to integrate more knowledge transfer using traditional university lecturing methods into the course: "\textit{No lecture part. I would have liked a four to six week lecture part in which we get some innovative PM methods taught by the prof.}"

\begin{table}[htbp]
\caption{Average scores on the adapted eduScrum approach and real projects}
\begin{center}
\begin{tabular}{p{5cm}|p{2cm}}
\hline
\textbf{Survey item} & \textbf{Average score} \\
\hline
My understanding of real world problems related to project management was promoted. & 1.4 \\
\hline
My interest on the course objectives and content was aroused.  & 1.9 \\
\hline
The importance of the material for my professional activity became clear to me. & 1.5 \\
\hline
Overall, I rate the didactic method (eduScrum) positively. & 1.6 \\
\hline
\end{tabular}
\label{tab-5}
\end{center}
\end{table}

There were also positive comments on the open formulated questions: "\textit{There was no personal interest in the topic, but it was implemented very well.}" Another student notes: "\textit{eduScrum is nice. Please continue with real projects.}"

\subsection{Lessons learned from virtual distance teaching during Covid 19}
The Covid 19 pandemic had a significant impact on everyday life around the world. These effects are also present in (German) universities since March 2020 \cite{Ng.2021}. Since then, face-to-face teaching has been primarily restricted and even prohibited during lock-down measures. The switch from a co-located on-site collaboration to a distributed remote activity poses various challenges for agile teams (e.g., \cite{Butt.2021, Marek.2021, Neumann.2021}). The challenges occur because many agile methods and practices are significantly influenced by collaboration and communication.

We had the opportunity to carry out the introduction and kickoff events during the first two weeks of class in person. This two weeks were important in order to get the chance to get to know each other personally.  For our role as agile coaches, the student teams, and the product owners, remote collaboration was still a challenge. The coordination was more complex, and the communication, especially in coaching and mentoring situations, did not reach the depth of content as in the previous year. Nonetheless, the switch to virtual distance teaching also resulted in some positive aspects. The coordination was often more spontaneous. The contact between the student teams and the product owners as well as agile coaches was also more besides the planned events, which, for example, led to quicker clarifications. We carried out the appointments with Skype for Business (and Zoom later in the summer term 2020). The connections were always stable, and there were no performance problems. This also applies to other tools such as Trello or cloud providers such as Dropbox.

We did not adapt the eduScrum approach in its methodical characteristics or how the practices are described. We attached more importance to the summer term 2020 because of the importance of the transparency of the Product and Sprint Backlog artifacts. This was not always optimal, which led to unnecessary expenditure of time at the beginning of the appointments.

What is still a significant challenge is building trust between the students and the agile coaches. The quality and intensity of face-to-face conversations in a class room could not be reproduced in virtual distance teaching. Retrospectives also never reached the depth they had in the previous year when they were conducted on-site and in person.

Since too many contextual factors are associated with the extensive switch to virtual distance teaching, we decided not to take the collected data into account in this study. From our point of view, the two semesters are not comparable in terms of implementation, especially due to the virtual distance teaching and other projects and companies, new stakeholders, and the new student groups.

\section{Limitations}
\label{s5:lim}
Several limitations should be taken into account in our study. First of all, we need to point out the small group of students in the course and the correlating small number of feedback from the survey (n = 21). We, therefore, used qualitative data collection methods to be able to verify the results of the survey. Our observations and informal discussions with the students and product owner confirmed the results of the survey. 

Another aspect that we have to consider as a limitation is when we conducted the survey. When the students took part in the survey, the final evaluation of their performance was not yet precise. This can lead to positive feedback from students and thus trigger bias. 

Also, we carried out the survey only once at the end instead of planning a second run for half of the term period. This would have enabled us to react to feedback in the semester and to carry out potential adaptations in the procedure. However, we did not receive any feedback or ideas in the discussions with the student teams we will take this as a reminder for our future work.

\section{Conclusion and Future work}
\label{s6:conc}
The first goal of our study was to integrate the agile education method eduScrum considering real world problems in form of projects. We adapted the eduScrum approach to our needs and used mainly characteristics of agile practices, artifacts and roles from well-known agile methods like Scrum. The adapted eduScrum approach was used in the summer terms 2019 and 2020 a Master of Science class dealing with innovative methods of agile project management at a German university. Our findings base on the data collected during summer term 2019 due to the Covid 19 impact on the teaching activities in 2020. We collected the data via a survey and qualitative methods like informal talks and discussions as well as observations during the eduScrum events.

The students value working with eduScrum in combination with real projects. We aimed to generate real experiences for the students while not decreasing the quality of the transfer of knowledge.  We also value the experience while working with eduScrum due to the changed role towards a coach. 

Based on our findings, we optimized our approach especially concerning the project goals and requirements. We currently performing a class using eduScrum with more academic focused projects for the student groups. The projects aim to plan, perform, analyze and report experiments with real agile software development teams on topics like influencing factors of remote work, e.g., on the team´s performance or the product´s quality. We are planning to report our results in a study in the near future.

\section*{Appendix A}
The survey is available at the academic cloud: https://sync.academiccloud.de/index.php/s/MHkYiBA3x4hOziA

\bibliographystyle{plain}
\bibliography{references_new}

\end{document}